\documentclass[preprint,showpacs,preprintnumbers,amsmath,amssymb]{revtex4}

\usepackage{graphicx}
\usepackage{dcolumn}
\usepackage{bm,ulem,color}
\usepackage{amsmath, amssymb, mathrsfs}

\newcommand{\ra}{\rightarrow}
\newcommand{\RR}{\mathbb{R}}

\newcommand{\n}{\boldsymbol{n}}
\newcommand{\bb}{\boldsymbol{b}}

\newcommand{\ud}{\mathrm{d}}

\newcommand{\M}{\rm M}

\newcommand{\m}{\rm m}

\begin{document}

\title{A continuous model for microtubule dynamics with catastrophe, rescue and
nucleation processes}
\author{Peter Hinow}
\email{hinow@ima.umn.edu }
\affiliation{ Institute for Mathematics and its Applications,
University of Minnesota, 114 Lind Hall, Minneapolis, MN 55455, USA.}
\author{Vahid Rezania}
\email{vrezania@phys.ualberta.ca}
\altaffiliation[Also at ]{Department of Science, Grant MacEwan College, Edmonton, AB T5J 2P2, Canada.}

\author{Jack A.~Tuszy\'nski}
\email{jtus@phys.ualberta.ca}
\affiliation{Department of Physics, University of Alberta, Edmonton, T6G 2J1, Canada.}

\date{\today}

\begin{abstract}
Microtubules are a major component of the cytoskeleton distinguished by highly dynamic behavior both
in vitro and in vivo referred to as dynamic instability. We propose a general mathematical model that accounts for the growth, catastrophe,
rescue and nucleation processes in the polymerization of microtubules from tubulin dimers. Our model is
an extension of various mathematical models developed earlier formulated in order to capture and unify the
various aspects of tubulin polymerization. While attempting to use a minimal number of adjustable
parameters, the proposed model covers a broad range of behaviors and has predictive features discussed in the paper. We have
analyzed the range of resultant dynamical behavior of the microtubules by changing each of the parameter values at a time and
observing the emergence of various dynamical regimes, that agree well with the previously reported experimental data.
observing the emergence of various dynamical regimes,  that agree well with reported experimental behavior. 
\end{abstract}


\maketitle

\section{Introduction}
Microtubules are protein polymers made of $\alpha/\beta$ tubulin
heterodimers that form an essential part of the cytoskeleton of all
eukaryotic cells. Besides giving structural stability and rigidity
to a cell, microtubules play key roles in many physiological
processes such as intracellular vesicle transport and chromosome
separation during mitosis. An individual microtubule is a hollow
cylinder of $25\,nm$ diameter built usually from 13 protofilaments
\cite{BLT99}. While the stable subunits of microtubules are actually
heterodimers composed of $\alpha$ and $\beta$ monomers, we will
refer to them for simplicity as monomers. These monomers exist in two different energetic states, namely bound to a molecule of guanosine triphosphate (GTP) or guanosine diphosphate (GDP), respectively. Only the GTP bound monomers are assembly competent meaning they are able to polymerize into microtubules. After the GTP monomers have been added to the growing microtubule, GTP bound to $\beta$ tubulin is rapidly hydrolyzed (dephosphorylated) to form a bound GDP subunit. It has been hypothesized since the early 1980s that the so-called GTP cap on the tip of the growing microtubule gives rise to the stability of the microtubule \cite{Mitchison}. Once the GTP cap is lost, the microtubule will switch to a ``collapsing'' state, referred to as a catastrophe which is characterized by rapid depolymerization of the microtubule into its free subunits. However, there is also a possibility that at some point in time a catastrophically shrinking microtubule acquires a new GTP cap and thereby returns to the growing population, a situation that is referred to in the literature as a rescue event. We refer the reader to the papers \cite{FHL94,FHL96, BLT99, BLT99b,Sept,Antal,Cassimeris,Ranjith} and the references therein for more information about this phenomenon known in the literature as \textit{dynamical instability}. The process of microtubule polymerization dynamics both in vitro and in vivo has been exhaustively reviewed by Desai and Mitchison \cite{Desai}.  It should be stressed that under the conditions of high concentration of tubulin a completely different process has been observed, namely a transition to a regime with damped oscillations of the tubulin mass polymerized into microtubules. This occurs for free tubulin concentrations exceeding a critical value, that depends on the experimental conditions and which corresponds to saturable polymerization kinetics \cite{Carlier,Marx}. Dynamic instability of microtubules has been observed in vivo and in vitro \cite{Mitchison} and highlights both the non-equilibrium nature of the physical problem involved and the stochastic nature of the individual microtubule growth process.  Interestingly, ensembles of  microtubules show collective oscillations under suitable conditions first reported by Carlier \textit{et al.}~\cite{Carlier,Marx} and further studied by Mandelkow \textit{et al.}~\cite{Mandelkow89,Marx,Marx94}  who showed that above a critical value of the free tubulin concentration, the amount of polymerized tubulin undergoes smooth oscillations which are damped out as the biochemical energy source in the form of GTP molecules is gradually depleted.  Sept \textit{et al.}~\cite{Sept}   modeled the assembly dynamics from a chemical reaction-kinetics standpoint and found good agreement with the experimental data. While this model captures the main features of the process, it is still very empirical and incomplete. Nonetheless, to introduce the main features of the underlying biochemical processes we have described them in Appendix \ref{App_B}. These processes involve polymerization/depolymerization, nucleation, and catastrophe events. When at least one auto-catalytic reaction is added to the system, the dynamics change significantly.  Sept \textit{et al.}~\cite{Sept}, for example, considered an induced catastrophe event described by and incorporated into their model to reproduce oscillations observed in vitro. We will show in our paper that oscillations can be reproduced by including growth and nucleation in the model as the only nonlinear reactions.

Deterministic mathematical models of polymer growth largely fall into two classes, depending on whether the length of the polymer is discrete or continuous. The latter approximation is based on the assumption that the typical length of a polymer is much larger than the gain in length by adding a single monomer unit. Partial differential equation models of this type have been used for example in \cite{FHL96,BLT99,Sept,Marx94,Bayley,Chen,Houchmandzadeh} for the case of microtubule dynamics and in \cite{Greer} to study the dynamics of prion proliferation.

\section{The mathematical model }
In this paper, we propose a mathematical model for the concomitant processes of microtubule growth, nucleation, catastrophic shrinking and rescue. Roughly speaking, microtubules of length $x$ grow at a velocity $\alpha p(t)$ where $\alpha$ is a constant and $p(t)$ the concentration of free GTP tubulin. The model will therefore have the form of a nonlinear transport equation. This is essentially also the approach taken in some of the earlier works in this area \cite{FHL94,FHL96,JWF97,BLT99}. Our intention is to develop a generalized model that accounts for the wealth of observed behavior including nucleation, growth to saturation and synchronized oscillations, in addition to catastrophes and rescues. Moreover, we have explicitly included the presence of a GTP cap and a measure of the microtubule's age. While striving for completeness in the mathematical description we have also attempted to introduce a minimal number of model parameters. Our model contains $7$ empirical parameters, most of which can be determined from experimental data.

\begin{figure}[th]
\begin{center}
\includegraphics[width=120mm]{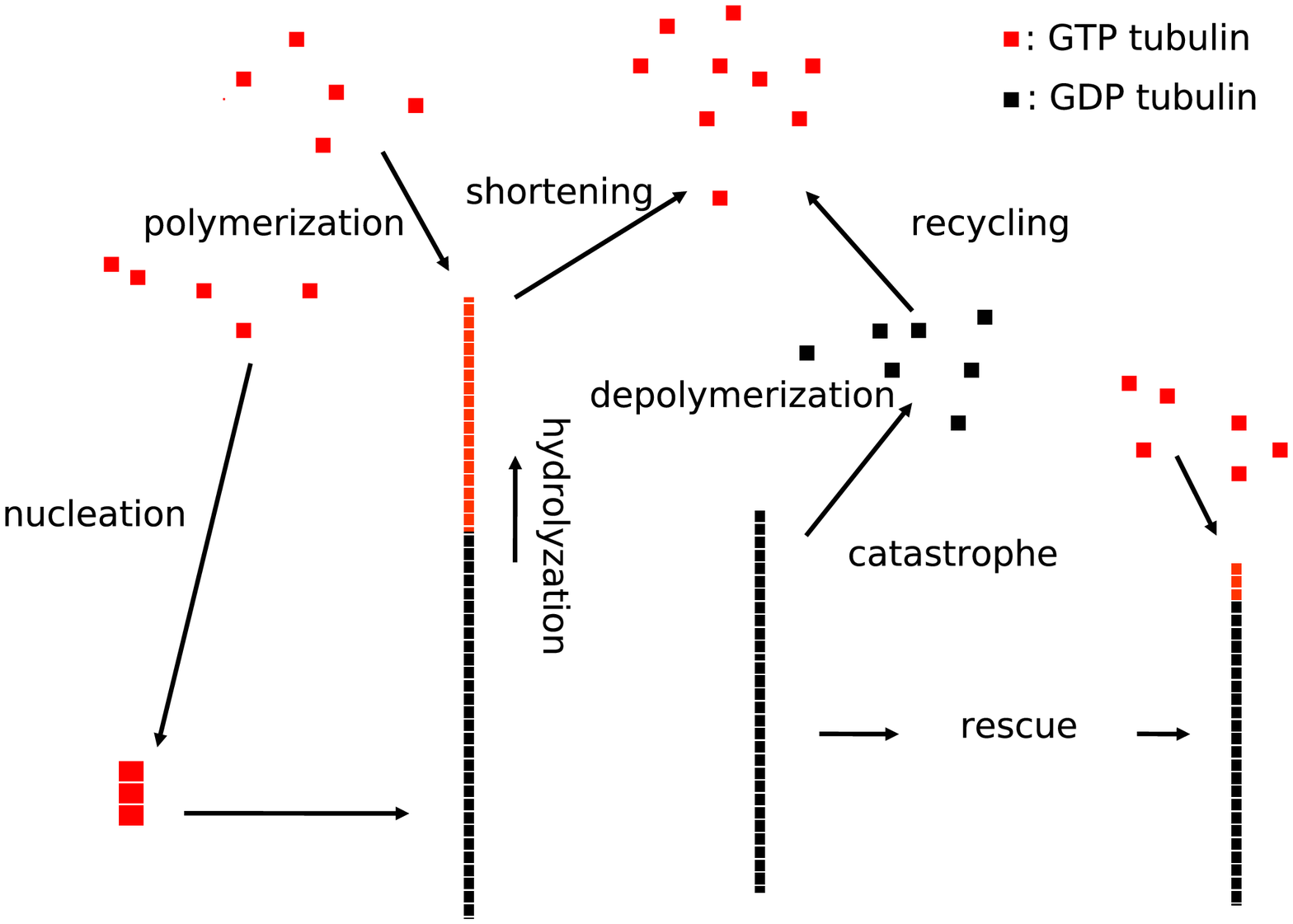}
\includegraphics[width=120mm]{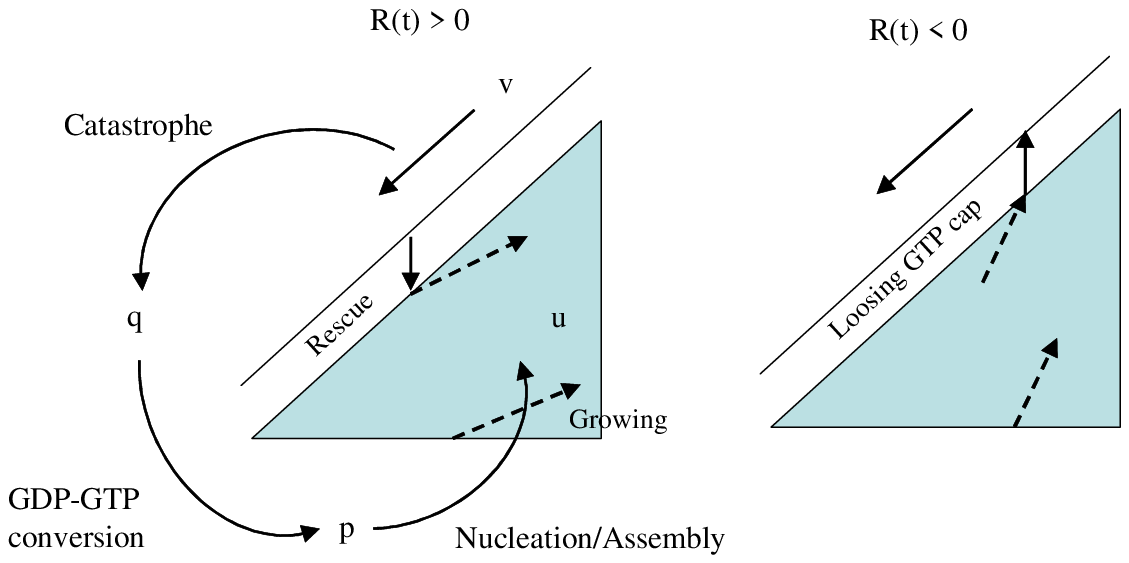}
\caption{Schematic representation of the tubulin cycle (top) and our model presented in this paper (bottom).
In the case  $R(t)= \alpha p(t)-\beta-\gamma >0$, the system  is in a phase of growth
(dashed arrows, left panel). If $R(t) <0$, the system  is in a phase of shrinking
(dashed arrows, right panel). The recycling of free monomers is identical in both
cases and only depicted in the left panel.}\label{Fig1}
\end{center}
\end{figure}

Let $Y=\{(x,y)\in\RR^2\::\:x> y> 0 \}$ be the state space for microtubules with a  GTP cap. 
For $(x,y)\in Y$, let $u(x,y,t)$ denote the population density of microtubules of total length $x$ that have a GTP cap and whose GDP domain has length $y$. This implies that
\begin{equation*}
||u(\,\cdot\,,\,\cdot\,,\,t)||_{\,\ud y\,\ud x} = \int_0^\infty
\int_0^x u(x,y,t)\,\ud y\,\ud x
\end{equation*}
is the total number of microtubules in a reference volume.  
We assume that the GDP domain forms a connected set and that the remainder of the microtubule is the GTP domain of length $x-y$, we will return to  this assumption in the discussion section below. If the microtubule has no GTP cap then it will undergo catastrophic depolymerization. In order to keep track of this process, let $v(x,t)$ denote the density of microtubules of length $x$ without a GTP domain, again in the sense of a concentration. In addition, we introduce the concentrations of free GTP monomers $p$ and free GDP monomers $q$. Catastrophic depolymerization of microtubules results in the release of GDP monomers which are then biochemically converted into GTP monomers in a reaction sometimes referred to as ``pumping'' since it involves a biochemical energy input from the solution. The new GTP monomers then become available for further microtubule growth, rescue and nucleation. Fig.~\ref{Fig1} provides a schematic depiction of the tubulin cycle. The equation for $u$ is given by 
\begin{equation}\label{model_B_eqn1}
\frac{\partial}{\partial t}u(x,y,t)+(\alpha p(t)-\beta)\frac{\partial}{\partial x}u(x,y,t)+\gamma\frac{\partial}{\partial y}u(x,y,t)=0.
\end{equation}
The new monomers are added at rate $\alpha p(t)$ and result in an increase of the overall lengths of the microtubules. The constant $\gamma>0$ is the progression rate of the GDP zone (i.e.~the speed of hydrolysis within the microtubule). The rate $\beta$ can be positive if occasionally a GTP bound monomer is lost from the microtubule. Notice that both factors $\alpha p-\beta$ and $\gamma$ have the dimension $LT^{-1}$. The characteristic curves for equation \eqref{model_B_eqn1} are given by
\begin{equation*}
\frac{dx}{dt} = \alpha p(t)-\beta, \qquad \frac{dy}{dt} =\gamma.
\end{equation*}
In view of the second of these equations, the variable $y$ can also be interpreted as the ``age'' of the microtubule, since hydrolysis is assumed to start immediately upon nucleation. The boundary condition on $\Gamma_1 = \{(x,y)\in\RR_{\ge0}^2\::\: y=0\}$ incorporates the nucleation of microtubules without a GDP domain. Let $\psi(x)$ be the length distribution of such freshly nucleated microtubules without a GDP domain. This can be a uniform distribution on some interval $[x_-,x_+]$ or a (narrow) Gaussian distribution centered at some point $x^0$.  Let $L^*=\int_0^\infty x \psi(x)\,\ud x$ be the average length of freshly nucleated microtubules. The nucleation reaction is generally assumed to be a nonlinear reaction although the exact number of monomers $n$ that need to come together is a matter of discussion \cite{BLT99b,Fygenson,Jackson}. With the rate of nucleation $\mu>0$, the boundary condition on $\Gamma_1$ is
\begin{equation}\label{model_B_eqn1b}
\gamma u(x,0,t) =  \frac{\mu}{L^*} p^n(t) \psi (x).
\end{equation}
Thus $\Gamma_1$ is part of the inflow boundary of the domain $Y$ at
all times.
The boundary $\Gamma_2=\{(x,y)\in\RR_{\ge0}^2\::\: x=y\}$ may be part of the inflow boundary respectively of the outflow boundary, depending on whether the growth of the entire microtubule is faster than the progression of the GDP domain. Precisely, let $R(t) = \alpha p(t)-\beta-\gamma$. If $R(t)>0$, then we say that the system is in a growth phase. This allows microtubules without a GTP cap to be rescued. The boundary condition on $\Gamma_2$ is then given by 
\begin{equation}\label{model_B_eqn1c}
R(t) u(x,x,t) = \lambda  v(x,t), \: \textrm{ if }\: R(t) >0,
\end{equation}
where $\lambda\ge 0$ is the propensity of shrinking microtubules to be rescued (it has the dimension $T^{-1}$). If on the other hand, $R(t) < 0$, then we say that the system is in a state of shrinking
and $\Gamma_2$ is part of the outflow boundary of the domain $Y$. Microtubules reaching the boundary $\Gamma_2$ are transferred to the population of microtubules without a GTP cap $v(x,t)$. It will be
helpful later to write equation \eqref{model_B_eqn1} in divergence form. Let
\begin{equation*}
\bb(t) = \left( \begin{array}{cc} \alpha p(t) -\beta \\ \gamma \end{array}  \right) ,
\end{equation*}
then equation \eqref{model_B_eqn1} can be written as
\begin{equation}\label{model_B_eqn1_divform}\tag{1'}
u_t+\nabla\cdot(\bb(t) u)=0.
\end{equation}

Microtubules without a GTP cap are shrinking at a rate $\delta>0$ (which represents the loss of length per unit time). This population has a source or a loss term, again depending on the sign of the
function $R$. The equation for $v$ is
\begin{equation}\label{model_B_eqn2}
\frac{\partial}{\partial t}v(x,t)-\delta \frac{\partial}{\partial x}v(x,t) = \left\{
      \begin{array}{ll}
      -R(t) u(x,x,t), & \textrm{ if } R(t)< 0 \\
      -\lambda v(x,t),   & \textrm{ if } R(t)>0
      \end{array}
  \right. .
\end{equation}
If $R(t)<0$ the system is in a state of shrinking, and microtubules that lose their GTP cap obviously enter the population of microtubules without a GTP cap. If the system is in a state of growth then microtubules without a GTP cap are rescued and re-enter the class $u$ through the boundary condition \eqref{model_B_eqn1c} on  $\Gamma_2$.

GDP-bound monomers are gained by catastrophes (there are no
intermediate depolymerization products) and are converted to
GTP-bound monomers by biochemical ``pumping''. The gain is
proportional to the number of microtubules without a GDP domain.
Hence
\begin{equation}\label{model_B_eqn3}
\frac{d}{dt}q =\delta \int_0^\infty v(x,t)\,\ud x - \kappa q.
\end{equation}
The constant $\kappa >0$ denotes the rate of the (first order)
pumping reaction. We assume that there is always enough GTP
available in the solution to ensure a constant rate of pumping, but
in the future we may also include free GTP as a variable into the model.  Recall that $q$ and $p$ are concentrations of ``lengths'' of microtubules stored in free GDP or
GTP-bound tubulin monomers, respectively. In order to obtain the
concentrations of molecules, one can calculate $\tilde{q}=\ell^{-1}
q$, where $\ell$ is length gained by adding a single monomer.
Flyvbjerg \textit{et al.} \cite{FHL94, FHL96} used a simple
conversion where $\ell = 8\,nm/13 = 0.6\,nm, $ since $8\,nm$ is the
length of a single tubulin heterodimer and $13$ is the number of
protofilaments in a microtubule. 

The population of free GTP-bound tubulin monomers is replenished by
the conversion of GDP monomers while losses occur due to growth and
nucleation of microtubules. Therefore,
\begin{equation}\label{model_B_eqn4}
\frac{d}{dt}p  = -(\alpha p-\beta)\int_0^\infty\int_0^x u(x,y,t)\,\ud y\,\ud x
+ \kappa q -  \mu p^n.
\end{equation}
The last term in equation \eqref{model_B_eqn4} indicates that $n$
individual monomers combine during nucleation.
The set of initial conditions is
\begin{equation*}
u(x,y,0) = u^0(x,y),\quad v(x,0) = v^0(x), \quad p(0) = p^0, \quad q(0) = q^0.
\end{equation*}

We calculate the total length $||u||_{x\,\ud y\,\ud x}$ of GDP and
GTP-bound tubulin found in microtubules with a GTP cap by
integrating with the weight  $x\,\ud y\,\ud x$,
\begin{equation*}
||u(\,\cdot\,,\,\cdot\,,\,t)||_{x\,\ud y\,\ud x} =
\int_0^\infty \int_0^x u(x,y,t)x\,\ud y\,\ud x.
\end{equation*}
The total length of GDP-bound tubulin in collapsing microtubules is
given by a similar expression
\begin{equation*}
 ||v(\,\cdot\,,\,t)||_{x\,\ud x} = \int_0^\infty v(x,t)x\,\ud x.
\end{equation*}
The model \eqref{model_B_eqn1}-\eqref{model_B_eqn4} conserves the
total length of bound and free tubulin $||u||_{x\,\ud y\,\ud
x}+||v||_{x\,\ud x}+q+p$, i.e.
\begin{equation}\label{total conservation}
\frac{d}{dt}\left(\frac{}{}||u(t)||_{x\,\ud y\,\ud
x}+||v(t)||_{x\,\ud x}+q(t)+p(t)\right)=0,
\end{equation}
see Appendix \ref{App_A} for details. It is also possible to calculate the
total length of GDP-bound tubulin found in microtubules with a GTP
cap as
\begin{equation}\label{bound-GDP}
\ell_{\rm bound-GDP}(t)= \int_0^\infty \int_0^x u(x,y,t)y\,\ud y\,\ud x,
\end{equation}
and so the complementary quantity, the total length of GTP-bound
tubulin, is
\begin{equation*}
\ell_{\rm bound-GTP}(t)= \int_0^\infty \int_0^x u(x,y,t)(x-y)\,\ud y\,\ud x.
\end{equation*}

\section{Parametrization and numerical results}

The parameters of our model are the following, each given with their
physical dimensionality
\begin{itemize}
\item $\alpha$, the growth rate of microtubules such that $\alpha p(t)$
has dimension $LT^{-1}$,
\item $\beta$, rate of loss of a GTP-bound monomer ($LT^{-1}$),
\item $\gamma$, progression rate of the GDP zone  ($LT^{-1}$),
\item $\delta$, depolymerization rate of microtubules without a GTP cap  ($LT^{-1}$),
\item $\kappa$, rate of the ``pumping'' reaction that converts GDP-bound monomers
into GTP-bound monomers ($T^{-1}$),
\item $\lambda$, rescue propensity for microtubules undergoing a catastrophe ($T^{-1}$),
\item $\mu$, rate of nucleation ($[p]^{-(n-1)} T^{-1}$),
\item $n$, the order of the nucleation reaction,
\item $\psi$, distribution of lengths of freshly nucleated microtubules,
\item $L^*$, average length of freshly nucleated microtubules ($L$).
\end{itemize}
The dimensions of $\alpha$ and $\mu$ require some discussion. From
equation \eqref{model_B_eqn4} we see that $[\alpha p] = [\beta] =
LT^{-1}$ and therefore $[\alpha] = [p]^{-1} LT^{-1}$. Hence, we are
only able to use sources that report the growth rate as dependent on
the concentration of free GTP-bound tubulin. Equation
\eqref{model_B_eqn4} implies further that $[\mu] =
[p]^{-(n-1)}T^{-1}$.
We collect in Table \ref{Tab1} a set of representative numerical
values for the parameters that have been published in the
literature.

We have implemented our model numerically using \textsc{matlab}, the
codes will be available from the authors upon request. To simulate
the system of equations \eqref{model_B_eqn1}-\eqref{model_B_eqn4},
we discretize $Y$ into $500 \times 500$ cells where each cell has a
dimension of $200\, n\m \times 200\, \textit{n}\m$. We use an upwind scheme
with adaptive time step for the partial differential equations and
the explicit Euler method for the ordinary differential equations
\cite{Quarteroni}. Since the amount of free GTP-tubulin $p(t)$ changes over
time, so does the growth velocity of the microtubules and hence an adaptive
choice of the time step is necessary to guarantee the Courant-Friedrichs-Lewy
condition \cite{Quarteroni}. We continuously keep track of the total amount of
tubulin in all its forms to guarantee that our numerical solution satisfies the
conservation law \eqref{total conservation}. Using the length of a single unit as stated
above, we can convert a concentration of tubulin as follows
\begin{equation*}
1\,\mu \M \equiv 3.76\cdot 10^{14} \,\frac{\mu \m}{L} =:C.
\end{equation*}

For consistency and comparison we choose most of the parameters from only two
experimental sources \cite{Jackson,Walker88}.  Walker \textit{et al.}~\cite{Walker88}
provide experimental estimates of the
polymerization rate $\alpha$, the loss rate of GDP monomers $\beta$
and the depolymerization rate $\delta$. The values (summed on both
plus and minus ends) are
\begin{equation*}
\begin{aligned}
\alpha&=2.5\,\mu \m\,\min{}^{-1}\,\mu \M^{-1}  \equiv 1.33 \cdot 10^{-15}
\,L\,\min{}^{-1}\:\left(=\frac{\alpha}{C}\right), \\
\beta&=2.4\, \mu \m\,\min{}^{-1}, \quad \delta = 50\, \mu \m\,\min{}^{-1}.
\end{aligned}
\end{equation*}
These authors also estimate the rescue and catastrophe rates as
$10\,\min^{-1}$ and $0.36\,\min^{-1}$ at $10 ~\mu \M$ tubulin
concentration, respectively. The latter are used to fine tune
the other parameters of our model, i.~e.~$\gamma,~ \kappa,$ and $ \lambda$. We note
that the integral on the boundary $\Gamma_2$ in
equation \eqref{number_of_mts} can serve as a definition of  the
catastrophe, respectively rescue frequency, depending on the sign of
$R$.  Therefore, we define the time-average rescue and catastrophe
rates from our model as follows
\begin{equation}\label{def_freq}
\begin{aligned}
k_{\rm cat} &= -\frac{1}{T}\int_0^T \int_{\Gamma_2} u(x,x,t)\,\ud x\, R(t)\,\ud t,
~~~ \textrm{ if } R(t) <0,\\
k_{\rm res} &= \frac{1}{T}\int_0^T \int_{\Gamma_2} u(x,x,t)\,\ud x\,R(t)\,\ud t,
~~~~~~\textrm{ if } R(t) >0,
\end{aligned}
\end{equation}
where $T$ is the total simulation time. For the nucleation reaction
we assume that $n=2$ and $\mu = 5.9 \cdot 10^{-3} \mu \M{}^{-1} \, \min{}^{-1}$
\cite[Table 1]{Jackson}.

As one initial distribution of microtubules with GTP cap we choose
\begin{equation}\label{standard_initial_datum}
u^0(x,y) =c \exp\left( -\frac{(x-10)^2}{5^2} -\frac{(y-5)^2}{2.5^2}\right),
\end{equation}
where the constant $c$ is chosen such that $||u^0||_{x\,\ud y\,\ud
x} \equiv 5\,\mu M$, this is half the concentration of the total
bound tubulin. The initial concentration of free GTP-bound tubulin
is $p^0 \equiv 5\,\mu M$. The remaining two initial data are chosen
to be $0$.

\begin{figure}[th]
\begin{center}
\includegraphics[width=200mm]{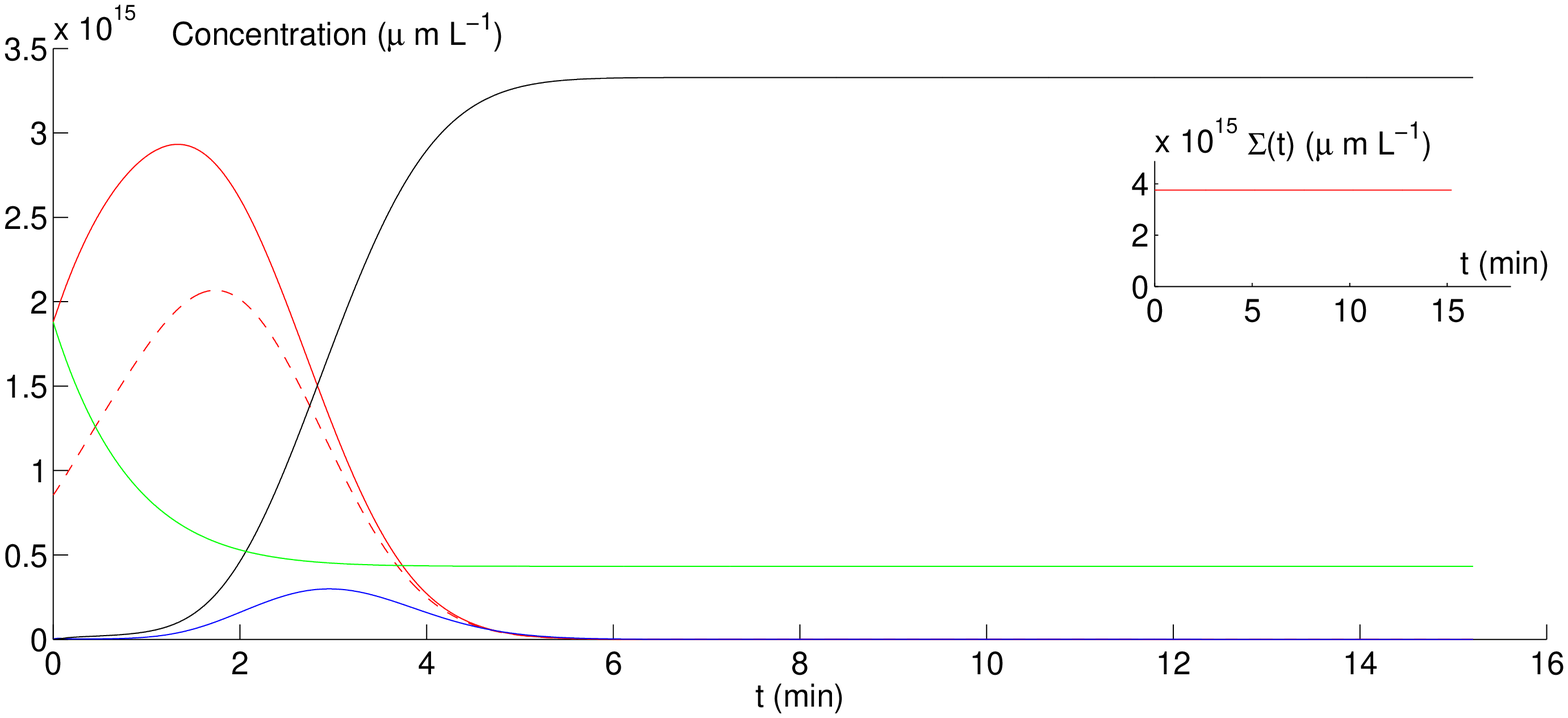}\vspace{1cm}\\
\caption{Time evolution of microtubules
for the parameter  set $\alpha= 2.5~\mu \m
\min^{-1} \mu \M^{-1}$, $\beta= 2.4\,\mu \m \min^{-1}$, $\gamma=
5.4~\mu \m \min^{-1}$, $\delta = 50~\mu \m\min^{-1}$ and $\mu =
\kappa = \lambda = 0$.
Shown are the total amounts of tubulin in each of its forms, namely bound
in microtubules with a GTP cap $||u||_{x\,\ud y\,\ud x}$ (solid red curve),
bound in microtubules without a GTP cap $||v||_{x\,\ud x}$ (blue curve), and
tubulin monomers bound to GTP $p$ (green curve) and bound to GDP $q$ (black curve).
All quantities are in units of $\mu \m L^{-1}$ of tubulin. The total length of
microtubules is decreasing at later times due to complete depolymerization of
microtubules without a GTP cap and lack of nucleation. The red dashed curve
represents the  amount of bound GDP tubulin in microtubules with GTP cap. The
inset plot shows the conservation  of the total amount of tubulin, confirming
equation \eqref{total conservation}.}
\label{Fig2}
\end{center}
\end{figure}

In the first modeling scenario we set the parameters as $\kappa=\lambda=\mu=0$ (no conversion of GDP monomers, no rescue and no nucleation), see Fig.~\ref{Fig2}. We plot the time evolution of
microtubules in $u$ (total amount of tubulin bound in microtubules with GTP cap, solid red curve) and $v$ (total amount of tubulin bound in microtubules without GTP cap, blue curve) pools,
and GTP and GDP tubulin dimers in $p$ (green curve) and $q$ (black curve) pools, respectively. After an initial period of growth, microtubules enter the depolymerization phase due to low concentration of the GTP tubulin (green curve).  This is because the GDP-GTP conversion process of monomers is turned off ($\kappa = 0$).  As a result, the total length of microtubules decreases. Due to lack
of nucleation and rescue ($\mu =0=\lambda$), this leads to complete depolymerization of all microtubules in the system, as expected. The  dashed red curve in  Fig.~\ref{Fig2}
represents the length of GDP-bound tubulin within microtubules with a GTP cap (see equation \ref{bound-GDP}).

\begin{figure}[th]
\begin{center}
\includegraphics[width=200mm]{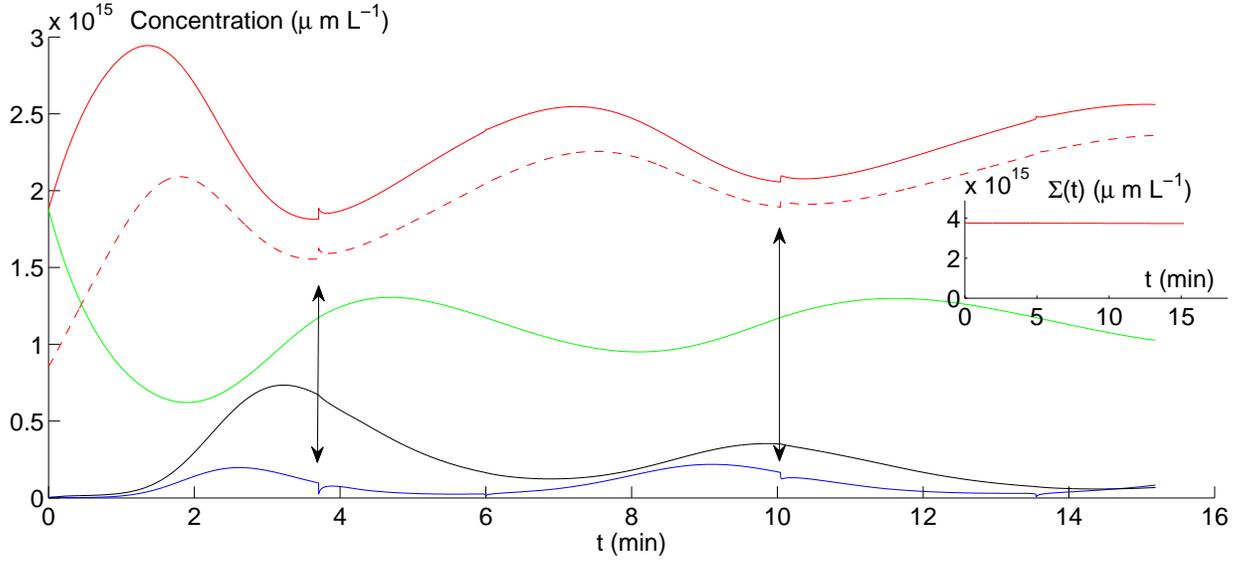}\vspace{1cm}\\
\caption{Same as Fig.~\ref{Fig2} but for the parameter set $\mu =5.9
\times 10^{-4} \mu \M{}^{-1} \, \min{}^{-1}$, $\kappa = 1\,\min^{-1}$ and $\lambda = 0.136\,\min^{-1}$. The arrows indicates a rescue events, the conservation of the total amount of tubulin is shown in the inset plot. The resulting catastrophe and rescue rates are $k_{\rm cat} = 0.1394~\min^{-1}$ and $k_{\rm res} = 6.7899~\min^{-1}$ (see equation \ref{def_freq}). The parameters $\alpha,\,\beta,\,\gamma$ and $\delta$ used in figure \ref{Fig2} and $\mu,\,\kappa$ and $\lambda$ used here are referred to as to the ``standard'' parameter set from now on.}\label{Fig3}
\end{center}
\end{figure}

\begin{figure}[th]
\begin{center}
\resizebox{14cm}{12cm}{\includegraphics[width=80mm]{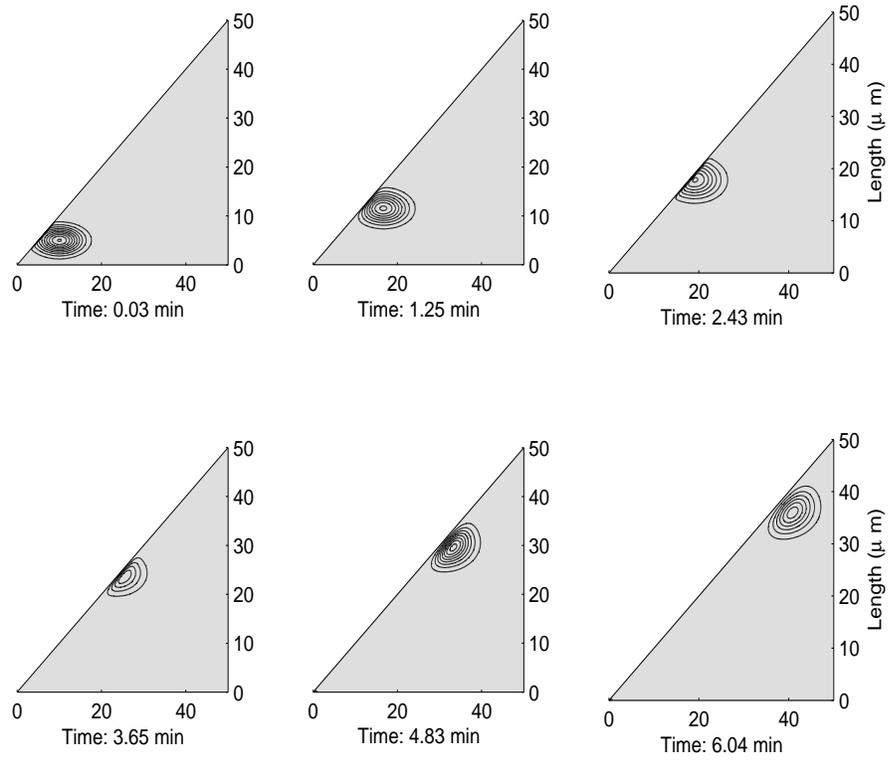}}
\caption{Time evolution of the population density $u$  for the set of parameters used in Fig. \ref{Fig3}. Plotted are equidensity contours beginning from a Gaussian profile. Note the change in direction of the center as time progresses due to the activation of the rescue process.
}\label{Fig4}
\end{center}
\end{figure}

\begin{figure}[th]
\begin{center}
\includegraphics[width=200mm]{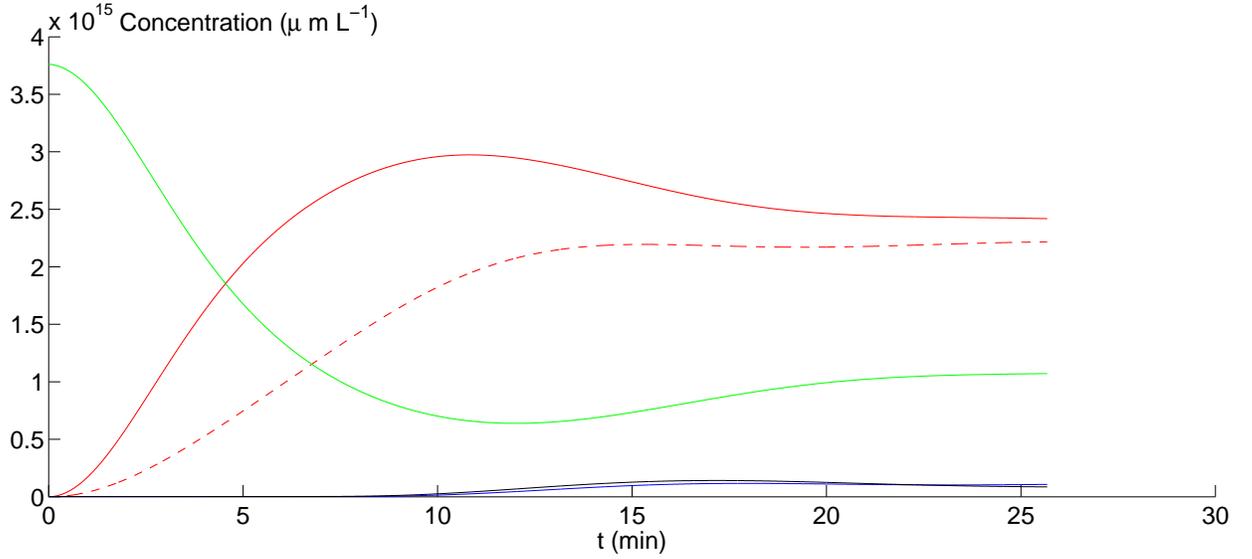}\vspace{1cm}\\
\caption{ Shown are again the total amounts of tubulin in each of its forms ($||u||_{x\,\ud y\,\ud x}$ solid red curve, $||v||_{x\,\ud x}$ blue curve, $p$ green curve and $q$ black curve). Parameters are
the same as in Fig.~\ref{Fig3}, but with initial conditions  $||u^0||_{x\,\ud y\,\ud x} = 0$ and $p^0= 10 \mu \M$.}\label{Fig5}
\end{center}
\end{figure}

In agreement with experimental observation \cite{Mandelkow89}, a solution showing damped oscillations can be found by introducing recycling of GDP tubulin, and nucleation and rescue processes. Using the parameters  $\kappa = 1\,\min^{-1}$, $\mu =5.9 \times
10^{-3} \mu \M{}^{-1} \, \min{}^{-1}$ and
$\lambda = 0.136\,\min^{-1}$, the result is shown in
Fig.~\ref{Fig3}.  From now on we refer to this parameter set
along with $\alpha=2.5\,\mu \m\,\min{}^{-1}\,\mu \M^{-1}$, $\beta=2.4\, \mu \m\,\min{}^{-1}$, and $\delta = 50\, \mu \m\,\min{}^{-1}$  as
to the ``standard'' and make all changes with respect to these values. In Fig.~\ref{Fig3}, kinks in the curves indicate rescue events, that occur when microtubules without a GTP cap acquire a GTP cap under favorable growth conditions. Setting the rescue rate $\lambda$ to zero results in very similar curves without the discontinuities (not shown). The resulting rescue and catastrophe rates are $k_{\rm res} = 6.7899\,\min^{-1}$ and $k_{\rm cat} = 0.1392\, \min^{-1}$ which are within experimentally observed ranges (see Table 1). Another interesting observation in Fig.~\ref{Fig3} is that the GDP zone (of the entire population) follows quite closely the total length. This means that on average, microtubules maintain their GTP cap. Figure \ref{Fig4} depicts the time evolution of the population density $u$ for the set of parameters described in Fig.~\ref{Fig3}. As shown, the microtubules continue to grow and shrink in time. Another interesting initial distribution of microtubules with GTP cap is $u^0(x,y) = 0$ and $p^0 \equiv 10\,\mu M$. In this scenario the influence of nucleation can be studied, see Fig.~\ref{Fig5}. As shown, microtubules are nucleating, polymerizing and then growing while the GDP-bound portion is also progressing.  

\begin{figure}[th]
\begin{center}
\includegraphics[width=140mm,height=65mm]{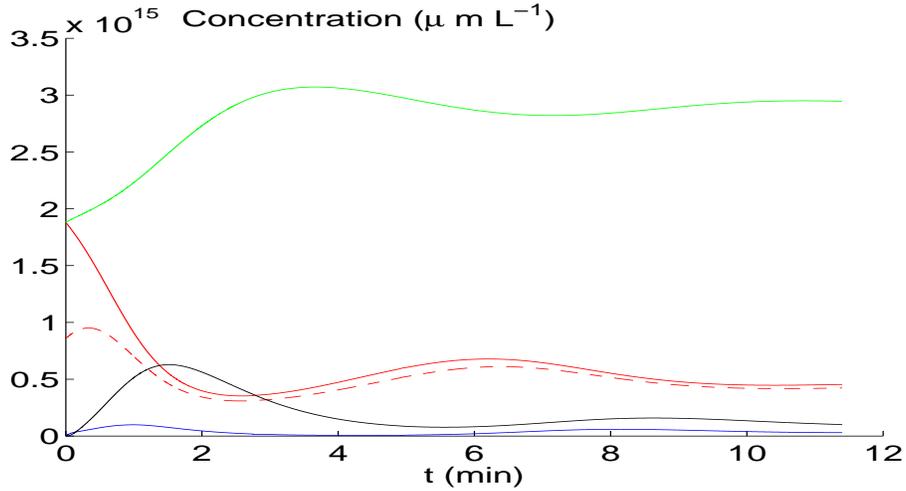}
\caption{An increase in the loss rate  of GTP  monomers $\beta$ to $6\beta^*$ does not prevent formation of microtubules entirely, but reduces drastically
their average length. Notice also the extremely short GTP cap. All other parameters are the same as in Fig.~\ref{Fig3}.}\label{beta_6}
\end{center}
\end{figure}

We have tested numerically the influence of the individual parameters on the behavior of our model. Using the initial datum from equation \eqref{standard_initial_datum}, we have varied one parameter at a time.In order to unify the picture somewhat, we arrange several growth curves in two-dimensional slices of the parameter space. In the following, we state variation of the relevant parameter as a multiple of the standard value, which is denoted by an asterisk. A decreased rate of monomer addition $\alpha \le0.3 \alpha^*$ results in no  growth of microtubules during 10 min  (Fig.~\ref{alpha_kappa}, left column). An increase in the  loss rate of GTP-bound monomers to  $6\beta^*$ results in the formation of much shorter microtubules with a very short GTP cap (Fig.~\ref{beta_6}). A reduced rate of GTP hydrolysis allows the formation of large GTP domains and delays the onset of the oscillation phase  (Fig.~\ref{alpha_gamma}, bottom row). On the other hand, large values of $\gamma$ are able to suppress microtubule formation completely (Fig.~\ref{alpha_gamma}, top row). It is worth noting that in Fig.~\ref{Fig3} we observe a crossover from a regime dominated by low-level microtubule assembly to saturation kinetics followed by a polymerization ``overshoot" with damped oscillations. The latter is a characteristic feature found experimentally by Marx \textit{et al.} \cite{Marx} almost 20 years ago.

\begin{figure}[th]
\begin{center}
\includegraphics[width=140mm,height=65mm]{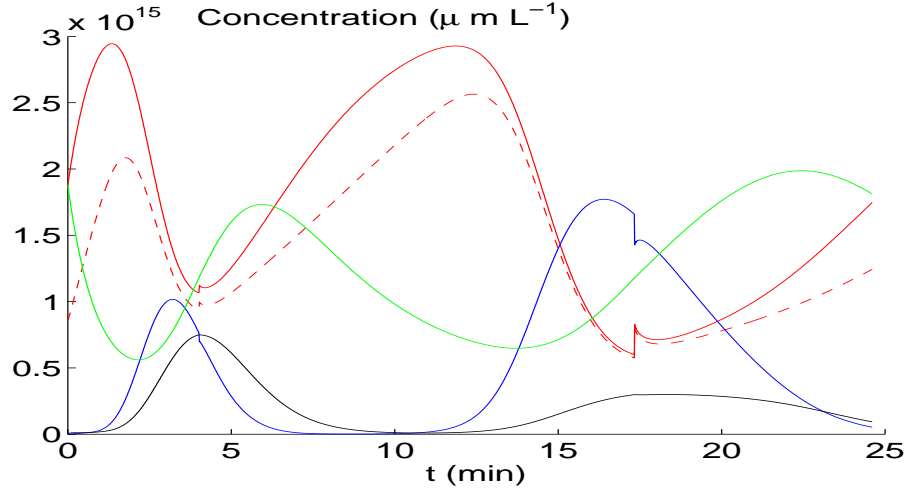}
\caption{A decrease of the depolymerization rate of collapsing microtubules to $0.2\delta^*$ results in irregular oscillations and longer presence of microtubules without a GTP cap. All other parameters are the same as in Fig.~\ref{Fig3}.}\label{delta_0_2}
\end{center}
\end{figure}

\begin{figure}[th]
\begin{center}
\includegraphics[width=170mm,height=85mm]{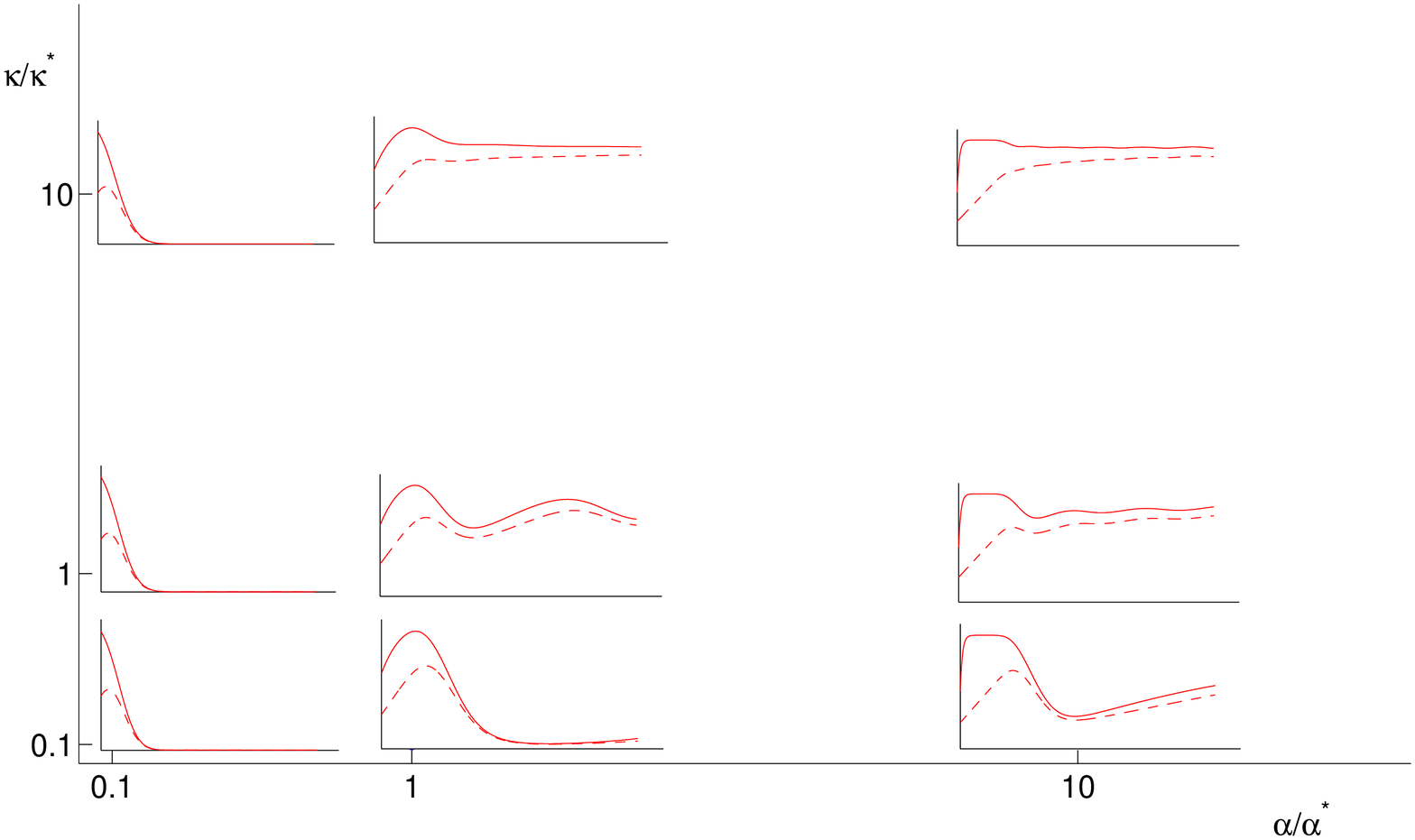}
\caption{The behavior at several points in parameter space when $\alpha$ and $\kappa$ are varied simultaneously. Each subfigure shows only the amount of tubulin in microtubules with GTP cap   (solid red curves) and the length of the trailing GDP subdomain (dashed red curves). Parameters are varied with respect to the values $\alpha^* = 2.5~\mu \m
\min^{-1} \mu \M^{-1}$ and $\kappa^* = 1\,\min^{-1}$. The time axis covers $10\,\min$ in each case. }\label{alpha_kappa}
\end{center}
\end{figure}

\begin{figure}[th]
\begin{center}
\includegraphics[width=170mm,height=85mm]{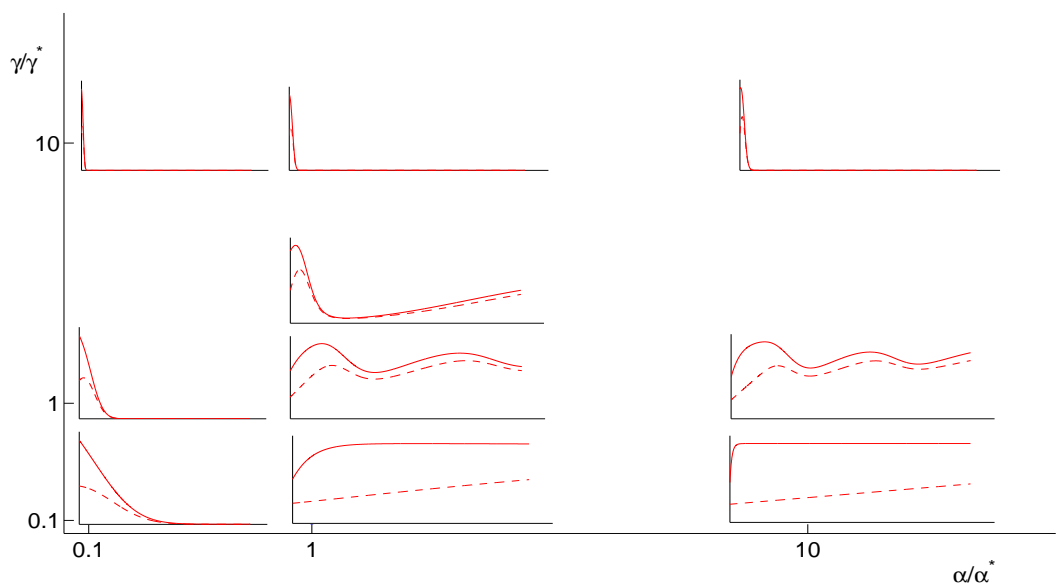}
\caption{As Fig.~\ref{alpha_kappa}, but now $\alpha$ and $\gamma$ are varied simultaneously.}\label{alpha_gamma}
\end{center}
\end{figure}

Changes of the depolymerization rate of microtubules without a GTP cap $\delta$ lead to interesting behavior. Strong oscillations and a prolonged existence of microtubules without GTP cap are seen (Fig.~\ref{delta_0_2}) when $\delta$ is small, and rescue events are more pronounced  due to a longer survival of  microtubules without a GTP cap. A very important role is played by the rate $\kappa$ at which GDP monomers are recycled to GTP monomers. Even a very small value such as $0.1\kappa^*$ suffices to induce oscillations, although their period is much longer (Fig.~\ref{alpha_kappa}, bottom row). On the other hand, a large value such as $10\kappa^*$ results in a quick dampening of the oscillations (Fig.~\ref{alpha_kappa}, top row). Finally, the parameter $\mu$ influences the timescale for the nucleation process, but does not change the general behavior drastically (not shown).

\begin{figure}[th]
\begin{center}
\includegraphics[width=140mm,height=65mm]{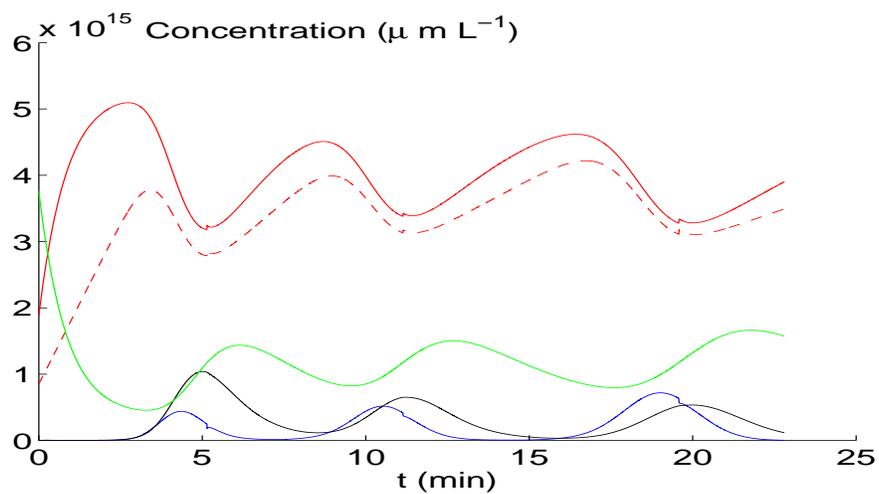}
\caption{Reverting to the standard parameter set (see Fig.~\ref{Fig3}) and using the initial condition $p^0\equiv 10\,\mu M$ gives oscillations that persist for longer times.}\label{p0_10}
\end{center}
\end{figure}

In addition to these variations of parameters we also varied one initial datum, namely the concentration of free GTP tubulin $p^0$. Now using the standard parameter set again, we see that a high concentration of $p^0 \equiv 10\,\mu M$ and $u^0(x,y)$ as in  equation \eqref{standard_initial_datum} results in a longer persistence of oscillations (Fig.~\ref{p0_10}). 
Our model provides results regarding the nature of the oscillatory  kinetics for microtubule assembly which is consistent with the original  experimental observations reported by Carlier \textit{et al.}~\cite{Carlier}[7], specifically the  data shown in their Fig.~1 where the amplitude of oscillations and their persistence diminsh as the concentration of free tubulin decreases from $150\, \mu\M$ to $50\, \mu\M$. The actual values of the tubulin concentration used in the experiment may not be directly comparable to our values due to the specific  experimental conditions under which the measurements were made.

\section{Discussion}
The  dynamical behavior of microtubules has attracted many investigators over the past few decades to examine the microtubule behavior in regard to many biophysical aspects \cite{FHL94,FHL96, BLT99, BLT99b,Sept,Antal, Ranjith}. In this paper, we have proposed a new mathematical model that includes all processes taking place during microtubule polymerization/depolymerization, namely: growth, nucleation, catastrophic shrinkage and rescue events. Our model contains the amounts of free tubulin in both its energetic forms (GTP and GDP-bound) as dependent variables. This results in a nonlinear transport equation whose mathematical analysis will pose a serious challenge. Nevertheless, we think this will give valuable insight into the role of the GTP cap in maintaining microtubule stability. The earlier paper by  Houchmandzadeh and Vallade \cite{Houchmandzadeh} contained expressions for free GTP and GDP tubulin, it did not contain a growth velocity for microtubules that was dependent on the amount of  free GTP  tubulin.  GTP monomers continue to be lost, even beyond the point when they have become completely depleted \cite[equation (9)]{Houchmandzadeh}. This results in possibly negative concentrations of  GTP monomers, avoided in our treatment.

A crucial assumption of our model is that the microtubules consist of two separated and connected domains, a GTP domain where the growth occurs and a ``trailing'' GDP domain. Only very recently, a paper by Dimitrov\textit{ et al.} \cite{Dimitrov} provided experimental evidence for the presence of a GTP cap in microtubules in vivo. The authors of \cite{Dimitrov} suggest further, that remnants of GTP tubulin left in the GDP domain play a role during rescue events and that growth resumes after such remnants have been exposed during a catastrophic depolymerization. This is clearly an exciting new development. Nevertheless, the remnants are likely to be very short (say one layer of GTP tubulin). It should be clarified that our model describes a large population of growing and shrinking microtubules where stochastic events in an individual microtubule have been averaged out.

We should also point out that our model does not contain a diffusion term of the type $div(D\nabla u)$ on the right hand side of equation \eqref{model_B_eqn1}. Such a term has been a prominent feature of earlier models \cite{FHL94, FHL96}, although it is not contained  in other models \cite{JWF97}. The role of ``length diffusion'' in a mathematical model for linear polymer accretion has been investigated from a mathematical point of view \cite{Collet,Laurencot}. Collet \textit{et al.} \cite{Collet} and Lauren\c cot and Mischler \cite{Laurencot} discussed convergence of the solutions of the discrete Becker-D\"oring system to solutions of the continuous Lifshitz-Slyozov equation under certain scaling  assumptions. The ``standard'' version of the Lifshitz-Slyozov equation in \cite{Collet,Laurencot} does not contain a diffusion term, implying that a population highly concentrated at a certain length initially does not disperse later on.

We have performed numerical simulations using mainly parameters from only two experimental sources, namely \cite{Jackson,Walker88}. Already with a few choices and variations of parameters, we are able to reproduce commonly seen dynamical behaviors, such as complete depolymerization in case of lacking recycling of GDP-monomers (see figure \ref{Fig2}) and damped oscillations in a growing population (figure \ref{Fig3}). The parameters that have not yet been determined experimentally are - to the best of our knowledge - the rescue rate $\lambda$ and the pumping rate $\kappa$, although the order of magnitude of $\kappa$ has been estimated by theoretical arguments \cite{BLT99b,Sept}. By varying these parameters in simulations we can predict their influence on the growth behavior and suggest experimental scenarios to look for. While outside of the scope of the present paper, we want to point out that microtubule polymerization and depolymerization is the target of many cancer chemotherapy drugs. The precise mechanism, by which some drugs (such as vinblastine and taxol) suppress dynamic instability is a topic for future modeling and experimental research. 

Some important points that we plan to address in the future are
\cite{Janosi08}
\begin{itemize}
\item The GTP zone is generally believed to be short, a few helical rings or $40\,nm$ at most \cite{Schek}.  This would imply that hydrolysis, under typical conditions, proceeds at roughly the same speed as the growth of the microtubules which in turn is dependent on the concentration of unpolymerized GTP-bound tubulin. Is hydrolysis of polymerized GTP-bound tubulin a catalyzed reaction? 
\item It is an open question whether the model developed in this paper can be adopted to describe situations corresponding to in vivo conditions, such as the presence of microtubule associated proteins (MAPs) during the polymerization process and the existence of discrete microtubule organizing centers. Some recent papers emphasize their role \cite{Brouhard,Vitre} in realistic representations of cellular processes.
\end{itemize}

\section*{\normalsize{Acknowledgments.}}
PH is supported by an IMA postdoctoral fellowship. Part of this research was carried out while PH was visiting the University of Alberta, Edmonton, and he wants to acknowledge the warm hospitality of this institution. Funding for this research at the University of Alberta has been provided by the Natural Sciences and Engineering Research Council of Canada (NSERC), the Canadian Space Agency (CSA) and the Alberta Cancer Foundation for which JAT expresses his gratitude. The authors would like to thank Imre J\'anosi for helpful discussions and Eric Carpenter for help with the numerical simulations. The careful reading of the manuscript and helpful comments of the referees greatly helped to improve the paper.

\appendix
\section{Balances for tubulin and microtubules}\label{App_A}
We integrate equation \eqref{model_B_eqn1_divform} over the domain $Y$ and apply the divergence theorem for weighted integrals 
\begin{equation*}
\int_\Omega(\nabla\cdot\bb)\, w(\xi)\,\ud\xi=
\int_{\partial\Omega}(\bb\cdot\n)\, w(\xi)\,\ud\sigma(\xi)
-\int_\Omega\nabla w\cdot\bb\,\ud\xi
\end{equation*}
where $\n$ is the outer normal vector. With $w(x,y)=x$ and $\n=\left(\begin{array}{c} 0\\-1\end{array} \right)$ on $\Gamma_1$ and  $\n=\left(\begin{array}{c} -1\\1\end{array} \right)$ on
$\Gamma_2$ this gives, interchanging the order of integration and
differentiation,
\begin{equation*}
\begin{aligned}
\frac{d}{dt}||u(t)||_{x\,\ud y\,\ud x} &=
\int_{\Gamma_1}\gamma u(x,0,t)x\,\ud x
-\int_{\Gamma_2}(-(\alpha p(t)-\beta)+\gamma) u(x,x,t)x\,\ud x \\
&\quad +\int_Y(\alpha p(t)-\beta)u(x,y,t)\,\ud y\,\ud x \\
&=\mu  p^n(t) + \left\{ \begin{array}{ll}
      R(t) \int_0^\infty u(x,x,t) x\,\ud x, & \textrm{ if } R(t)< 0 \\
      \lambda\int_0^\infty v(x,t) x\,\ud x, & \textrm{ if } R(t)>0
      \end{array} \right\} \\
&\quad + (\alpha p(t)-\beta)\int_0^\infty \int_0^x u(x,y,t)\,\ud y\,\ud x.
\end{aligned}
\end{equation*}
Likewise, we integrate equation \eqref{model_B_eqn2} with weight
$x\,\ud x$. Again, after integration by parts, we have
\begin{equation*}
\frac{d}{dt}||v(t)||_{x\,\ud x} = -\left\{ \begin{array}{ll}
      R(t) \int_0^\infty u(x,x,t) x\,\ud x, & \textrm{ if } R(t)< 0 \\
      \lambda\int_0^\infty v(x,t)   x\,\ud x, & \textrm{ if } R(t)>0
      \end{array} \right\} - \delta \int_0^\infty v(x,t)\,\ud x.
\end{equation*}
Adding these two results and equations \eqref{model_B_eqn3} and
\eqref{model_B_eqn4} yields equation \eqref{total conservation}.

If the weight $x$ is removed then we obtain the total number of
microtubules with or without a GTP cap. For microtubules with GTP
cap we obtain
\begin{equation}\label{number_of_mts}
\begin{aligned}
\frac{d}{dt}||u(t)||_{\ud y\,\ud x} &= \int_{\Gamma_1}\gamma u(x,0,t)\,\ud x
+(\alpha p(t) -\beta-\gamma)\int_{\Gamma_2} u(x,x,t)\,\ud x \\
&=\mu p^n(t) \int_{\Gamma_1} \psi(x)\,\ud x + R(t)\int_{\Gamma_2} u(x,x,t)\,\ud x,
\end{aligned}
\end{equation}
that is, such microtubules are gained through nucleation and lost or
gained (depending on the sign of $R(t)$) through exchange with the
population without GTP cap. For the latter we have
\begin{equation*}
\frac{d}{dt}||v(t)||_{\ud x} = -\delta v(0,t) - R(t)\int_{\Gamma_2} u(x,x,t)\,\ud x,
\end{equation*}
i.e.~microtubules without GTP cap are lost by complete
depolymerization and gained or lost through exchange with the
population with GTP cap. Taken together
\begin{equation*}
\frac{d}{dt}(||u(t)||_{\ud y\,\ud x} +||v(t)||_{\ud x}) =\mu p^n(t) \int_{\Gamma_1}
 \psi(x)\,\ud x -\delta v(0,t),
\end{equation*}
that is, only nucleation and complete depolymerization change the
total number of microtubules.

\section{Biochemistry of tubulin polymerization}\label{App_B}

The principal elements in the model of Sept \textit{et al.}~\cite{Sept} can be summarized by the equations that follow. For simplicity, the microtubule is considered as a linear polymer rather than an object of 13 protofilaments.  We shall denote a microtubule of $n$ subunits by $MT_n$. Note that in solution, the free tubulin subunits may be bound to either GTP or GDP at their exchangeable nucleotide site and we denote them as $T_{GTP}$ and $T_{GDP}$, respectively. Note that only tubulin bound to GTP is able to polymerize.  

The  reaction set involved in the process consists of addition, nucleation and catastrophic collapse is given by
\begin{equation*}
\begin{aligned}
MT_n+T_{GTP}&\leftrightharpoons MT_{n+1}, \\
n\,T_{GTP}&\stackrel{k_n}{\ra} MT_n, \quad \text{and}\\
MT_n &\stackrel{k_c}{\ra} n\,T_{GDP}.
\end{aligned}
\end{equation*}
For simplicity, it is assumed here that all collapses are complete and the number of dimers, $n'$  required for nucleation is an adjustable parameter not available directly from experiment. These equations can also be supplemented by the reactivation of tubulin, to make it assembly competent, which will occur when the concentration of GTP is high
\begin{equation*}
T_{GDP}+GTP\stackrel{k_r}{\ra}T_{GTP}+GDP.
\end{equation*}
The free energy change associated with this reaction is less than the free energy change of GTP hydrolysis in solution and the difference is attributed to a structural change in the tubulin dimer.  It is this conformational change which presumably makes assembly possible. 

\bibliography{microtubule_July6_PRE}
\newpage

\bf{Tables and Figures}

\begin{table}[th]
\begin{center}
\begin{minipage}{\textwidth}
\begin{tabular}{|c|c|c|}\hline\hline
parameter & value & reference \\
\hline
$\alpha$ &  $0.5 - 11.5  \,\mu \m\,\min^{-1}~\mu\M^{-1}$ &
\cite{Walker88,Gild92,SW93,CFK95,PW02}  \\
$\beta$ &  $1.6 - 35  \,\mu \m\,\min^{-1}$ &
\cite{Walker88,Gild92,SW93,CFK95,BLT99,Arnal00,Rusan,PW02}   \\
$\gamma$ &  $0.25  \,\mu \m\,\min^{-1}$ & \cite{FHL96} \\
$\delta$ &  $44 - 50  \,\mu \m\,\min^{-1}$ & \cite{Walker88} \\
$n$      &  $1-12$ & \cite{BLT99b,Sept,Fygenson,Jackson}  \\ 
$\mu$ &  $5.9\times10^3\, \M^{-1}\min^{-1}$ & \cite{Jackson}  \\
$\kappa$ &  $3-120\, \min^{-1}$ & \cite{BLT99b,Sept}\\
$k_{\rm res}$ &  $2-10\, \min^{-1}$ & \cite{Walker88,Rezania}\\
$k_{\rm cat}$ &  $0.1-1\, \min^{-1}$ & \cite{Walker88,Rezania}\\
\hline\hline
\end{tabular}
\end{minipage}
\caption{Experimental and/or computational estimates for parameters published
in the literature, some of which are
used in the model. $k_{\rm res}$ and $k_{\rm cat}$ are rescue and
catastrophe rates that can be used to fine tune $\lambda$, $\kappa$
and $\gamma$. }\label{Tab1}
\end{center}
\end{table}


%

\end{document}